\documentclass[preprints,article,accept,moreauthors,pdftex]{Definitions/mdpi} 

\firstpage{1} 
\pubvolume{1}
\issuenum{1}
\articlenumber{0}
\pubyear{2021}
\copyrightyear{2020}
\datereceived{} 
\dateaccepted{} 
\datepublished{} 
\hreflink{https://doi.org/} % If needed use \linebreak
\usepackage{graphicx}
\usepackage{caption}
\usepackage{subcaption}
\Title{Fully Automatic Deep Learning Framework for Pancreatic Ductal Adenocarcinoma Detection on Computed Tomography}

\TitleCitation{Fully Automatic Deep Learning Framework for Pancreatic Ductal Adenocarcinoma Detection on Computed Tomography}

\Author{Natália Alves $^{1}$*\orcidA{}, Megan Schuurmans$^{1}$, Geke Litjens $^{2}$, Joeran S. Bosma $^{1}$*\orcidB{}, John Hermans $^{2}$, Henkjan Huisman $^{1}$}

\AuthorNames{Natália Alves, Megan Schuurmans, Geke Litjens, Joeran S. Bosma, John Hermans and Henkjan Huisman}

\AuthorCitation{Alves, N.; Schuurmans, M.; Litjens, G.; Bosma, J.; Hermans, J.; Huisman, H.}
\address{%
$^{1}$ \quad Diagnostic Image Analysis Group, Department of Medical Imaging, Radboud University Medical Center, Nijmegen, The Netherlands

$^{2}$ \quad Department of Medical Imaging, Radboud Institute for Health Sciences, Nijmegen, The Netherlands}

\corres{Correspondence: natalia.alves@radboudumc.nl}

\abstract{Early detection improves prognosis in pancreatic ductal adenocarcinoma (PDAC) but is challenging as lesions are often small and poorly defined on contrast-enhanced computed tomography scans (CE-CT). Deep learning can facilitate PDAC diagnosis, however current models still fail to identify small (<2cm) lesions. In this study, state-of-the-art deep learning models were used to develop an automatic framework for PDAC detection, focusing on small lesions. Additionally, the impact of integrating surrounding anatomy was investigated. CE-CT scans from a cohort of 119 pathology-proven PDAC patients and a cohort of 123 patients without PDAC were used to train a nnUnet for automatic lesion detection and segmentation (\textit{nnUnet\_T}). Two additional nnUnets were trained to investigate the impact of anatomy integration: (1) segmenting the pancreas and tumor (\textit{nnUnet\_TP}), (2) segmenting the pancreas, tumor, and multiple surrounding anatomical structures (\textit{nnUnet\_MS}). An external, publicly available test set was used to compare the performance of the three networks. The \textit{nnUnet\_MS} achieved the best performance, with an area under the receiver operating characteristic curve of 0.91 for the whole test set and 0.88 for tumors <2cm, showing that state-of-the-art deep learning can detect small PDAC and benefits from anatomy information.}

\keyword{pancreatic ductal adenocarcinoma; deep-learning; early detection} 

\begin{document}
%%%%%%%%%%%%%%%%%%%%%%%%%%%%%%%%%%%%%%%%%%
\section{Introduction}
  Pancreatic ductal adenocarcinoma (PDAC) is the most common form of pancreatic cancer, which has the worst prognosis of all cancer diseases worldwide with a 5-year relative survival rate of only 10.8\% \cite{Ryan2014PancreaticAdenocarcinoma, CancerStatFacts-PancreaticCancerAccessed19/11/2021}. The incidence of pancreatic cancer is increasing, and it is estimated to become the second leading cause of cancer-related deaths in Western societies by 2030 \cite{CancerStatFacts-PancreaticCancerAccessed19/11/2021, Siegel2020Cancer2020}. Patients diagnosed in early disease stages, where the tumors are small (size<2cm) and frequently resectable, present a much higher 3-year survival rate (82\%) than patients diagnosed in later disease stages where the tumors are larger (17\%) \cite{Ardengh2003PancreaticResectability}. Unfortunately, tumors are rarely found in early stages and approximately 80–85\% of patients present with either unresectable or metastatic disease at the time of diagnosis \cite{Ryan2014PancreaticAdenocarcinoma}. Given these statistics, it is clear that early diagnosis of PDAC is crucial to improve patient outcomes, as reversing the stage distribution would more than double the overall survival, without any additional improvements in therapy \cite{Kenner2021ArtificialCancer}.

Early PDAC detection is challenging, as most patients do not present specific symptoms until advanced disease stages, and screening the general population is cost-prohibitive with current technology \cite{Gheorghe2020EarlySurvival, Kenner2021ArtificialCancer}. Furthermore, PDAC tumors are difficult to visualize in computed tomography (CT) scans, which are the most used modality for initial diagnosis, as lesions present irregular contours and poorly-defined margins \cite{Kenner2021ArtificialCancer}. This becomes an even more significant challenge in the initial disease stages as lesions are not only small (<2cm) but also often iso-attenuating, making them easily overlooked even by experienced radiologists \cite{HoYoon2011Small}. A recent study that reconstructed the progression of CT changes in prediagnostic PDAC, showed that suspicious changes could be retrospectively observed 18 to 12 months before clinical PDAC diagnosis. However, the radiologists' sensitivity at identifying those changes, and consequently referring patients for further investigation, was only 44\% \cite{Singh2020ComputerizedStudy}.

Artificial intelligence (AI) can potentially assist radiologists in early PDAC detection by leveraging high amounts of imaging data. Deep learning models, and more specifically convolutional neural networks (CNNs), are a class of AI algorithms especially suited for image analysis and have shown high accuracy in the image-based diagnosis of various types of cancer \cite{Esteva2017Dermatologist-levelNetworks, McKinney2020InternationalScreening, Yasaka2018DeepStudy}. CNNs take the scan as input and automatically extract relevant features for the diagnostic task by performing a series of sequential convolution and pooling operations.

Clinically relevant computer-aided diagnostic systems should have the ability to both detect the presence of cancer and, in the positive cases, localize the lesion in the input image, with minimal to none required user interaction. 

Recently deep learning models have started to be investigated for automatic PDAC diagnosis \cite{Zhu2019Multi-scaleAdenocarcinoma, Xia2020DetectingEnsemble, Ma2020ConstructionDiagnosis, Liu2020DeepValidation, K2021FullyTumors, Wang2021LearningPrediction}. However, most studies perform only binary classification of the input image as cancerous or not cancerous, without simultaneous lesion localization. Furthermore, the majority of publications do not focus on small, early-stage lesions, with only one study reporting the model performance for tumors with size < 2cm \cite{Liu2020DeepValidation}.

In this study, we hypothesize that state-of-the-art deep learning architectures can be used to detect and localize PDAC lesions accurately, especially regarding the subgroup of tumors with size < 2 cm. We propose a fully automatic deep-learning framework that takes an abdominal CE-CT scan as input and produces a tumor likelihood score and a likelihood map as output. Furthermore, we assess the impact of surrounding anatomy integration, which is known to be relevant for clinical diagnosis \cite{HoYoon2011Small}, on the performance of the deep-learning models. The framework performance is validated using an external, publicly available test set, and the results on the subgroup of tumors with size < 2cm are also reported.

%%%%%%%%%%%%%%%%%%%%%%%%%%%%%%%%%%%%%%%%%%
\section{Materials and Methods}
\subsection{Dataset}

This study was approved by the institutional review board (Radboud University Medical Centre, Nijmegen, The Netherlands), and informed consent from individual patients was waived due to its retrospective design. CE-CT scans in the portal venous phase from 119 patients with pathology-proven PDAC in the pancreatic head (PDAC cohort) and 123 patients with normal pancreas (non-PDAC cohort), acquired between January 1st, 2013 and June 1st, 2020, were selected for model development. 

Two publicly available abdominal CE-CT datasets containing scans in the portal venous phase were combined and used for model testing: (1) "The  Medical Segmentation Decathlon" dataset (MSD) from Memorial Sloan Kettering Cancer Center (USA), consisting of 281 patients with pancreatic malignancies \cite{Simpson2019AAlgorithms} , and (2) "The Cancer Imaging Archive" dataset from the US National Institutes of Health Clinical Center, containing 80 patients with normal pancreas \cite{Clark2013TheRepository, Pancreas-CTTheWiki} . 

The size of the tumors was measured from the tumor segmentation as the maximum diameter in the axial plane.

\subsection{Image Acquisition and Labeling}
The CE-CT scans were acquired with five scanners (Aquilion One, Toshiba [Tochigi, Japan]; Sensation  64  and SOMATOM  Definition  AS+,  Siemens  Healthcare [Forchheim, Germany]; Brilliance 64, Philips Healthcare [Best, Netherlands]; BrightSpeed, GE  Medical  system,  [Milwaukee,  WI,  USA]). The slice thickness was 1.0–5.0 mm, and  image  size  was either  512×512 pixels (232 images) or 1024x1024 pixels (10 images). Images with size 1024x1024 pixels were resampled to 512x512 prior to inclusion in model development. 

All images from the PDAC-cohort were manually segmented using ITK-SNAP version 3.8.0 by trained medical students, being verified and corrected by an abdominal radiologist with 17 years of experience in pancreatic radiology.
The annotations included the segmentation of the tumor, pancreas parenchyma, and six surrounding relevant anatomical structures, namely the surrounding veins (portal vein, superior mesenteric vein, and splenic vein), arteries (aorta, superior mesenteric artery, celiac trunk, hepatic artery, and splenic artery), pancreatic duct, common bile duct, pancreatic cysts (if present) and portomeseneric vein thrombosis (if present). 

\subsection{Automatic PDAC Detection Framework}
This study uses a segmentation-oriented approach for automatic PDAC detection and localization, where each voxel in the image is assigned either a tumor or non-tumor label. The models in the proposed pipeline were developed using the state-of-the-art, self-configuring framework for medical segmentation \textit{nnUnet} \cite{Isensee2021NnU-Net:Segmentation}. All models employed a 3D U-Net \cite{Cicek20163DAnnotation} as base architecture and were trained for 250,000 training steps with 5-fold cross-validation.

Regions of interest (ROIs) around the pancreas were manually extracted for both the PDAC and non-PDAC cohorts. An anatomy segmentation network was trained to segment the pancreas and the other anatomical structures (refer to the previous section), using the extracted ROIs from the scans in the PDAC cohort. This network was used to automatically annotate the ROIs from the non-PDAC cohort, which were then combined with the manually annotated PDAC cohort to train three different \textit{nnUnet} models for PDAC detection and localization: (1) segmenting only the tumor (\textit{nnUnet\_T}), (2) segmenting the tumor and pancreas (\textit{nnUnet\_TP}), (3) segmenting the tumor, pancreas and the multiple surrounding anatomical structures (\textit{nnUnet\_MS}). These networks were trained with two different initializations and identical 5-fold cross-validation splits, originating ten models for each configuration. The cross-entropy (CE) loss function was used for the PDAC-detection networks since it has been shown to be more suitable for segmentation-oriented detection tasks than the soft DICE+CE loss function, which is selected by default in the \textit{nnUnet} framework \cite{Baumgartner2021NnDetection:Detection, Saha2021AnatomicalFunctions}. Additionally, the full CE-CT scans from the PDAC cohort were downsampled to a resolution of 256$\times$256 and used to train a low-resolution pancreas segmentation network, which was then employed to automatically extract the pancreas ROI from unseen images during inference.

At inference time, images were downsampled, and the low-resolution pancreas segmentation network was used to obtain a coarse segmentation of the pancreas. This coarse mask was upsampled back to the original image resolution and dilated with a spherical kernel to close any existing gaps. Finally, a fixed margin was applied to automatically extract the ROI, which was the input to the previously described PDAC detection models. This extraction margin was defined based on the cross-validation results obtained with the PDAC cohort so that no relevant information is lost while cropping the ROI.

Each of the PDAC detection models (\textit{nnUnet\_T}, \textit{nnUnet\_TP} and \textit{nnUnet\_MS}) outputs a voxel-level tumor likelihood map, which indicates the regions of the image where the network predicts a PDAC lesion and the respective prediction confidence. In the case of the \textit{nnUnet\_TP} and \textit{nnUnet\_MS} networks, a segmentation of the pancreas is also produced. This segmentation was used in post-processing to reduce false positives outside the pancreas by masking the tumor confidence maps so that only the PDAC predictions in the pancreas region are maintained.

After post-processing, candidate PDAC lesions were extracted iteratively from the tumor likelihood map by selecting the voxel with maximum predicted likelihood and including all connected voxels (in 3D) with at least 40\% of this peak likelihood value. Then, the candidate lesion was removed from the model prediction, and the process was repeated until no candidates remained or a maximum of 5 lesions were extracted. The final output of the framework was a tumor likelihood defined as the maximum value of the tumor likelihood map. 

\begin{figure}[H]
    \centering
    \includegraphics[width=16cm]{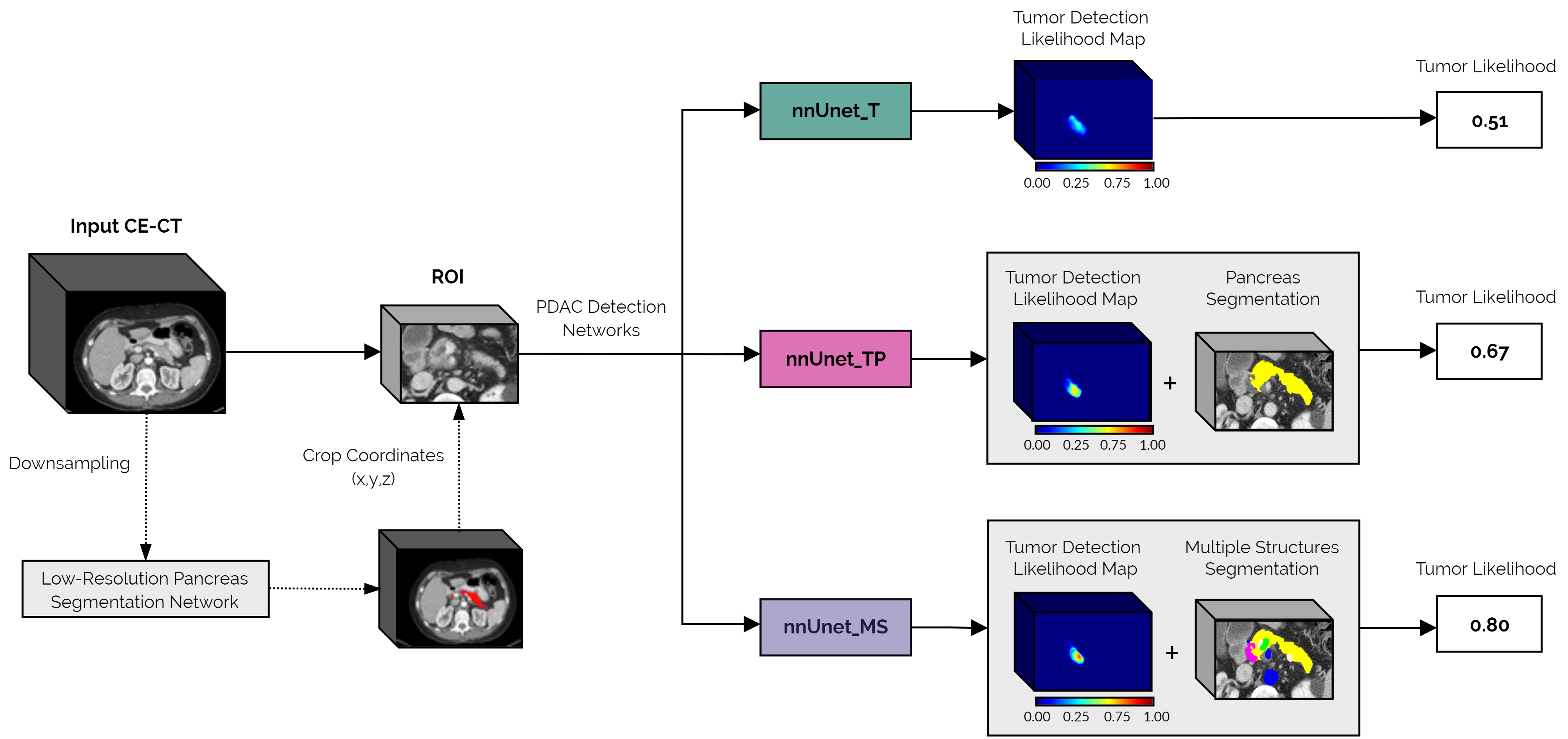}
    \caption{Schematic overview of the proposed automatic PDAC detection framework. The first step in the pipeline is to automatically extract the ROI from the full input CE-CT scan, using the low-resolution pancreas segmentation network. This ROI is then fed to each of the PDAC detection networks: \textit{nnUnet\_T}, \textit{nnUnet\_TP} and \textit{nnUnet\_MS}. The final tumor likelihood output is derived from the networks' tumor detection likelihood maps, which in the case of the \textit{nnUnet\_TP} and \textit{nnUnet\_MS} models is post-processed using the automatically generated pancreas segmentation. }
    \label{fig:inference_pipeline}
\end{figure}

 A schematic representation of the inference pipeline from the original image input to the final tumor likelihood prediction is shown in Figure~\ref{fig:inference_pipeline}.

\subsection{Analysis}
Patient-level performance was evaluated using the receiver operating characteristic (ROC) curve, while lesion-level performance was evaluated using the free-response receiver operating characteristic (FROC) curve. The ROC analysis assesses the model's confidence that a tumor is or is not present by plotting the true positive rate (sensitivity) against the false positive rate (1-specificity) at different thresholds for the model output, defined as the maximum value of the tumor likelihood map. The FROC analysis additionally assesses whether the model identified the lesion in the correct location, by plotting the true positive rate against the average number of false positives per image, at different thresholds for each individual lesion prediction \cite{Chakrabortv1989MaximumData, Bunch1977APerformance}. Each candidate lesion extracted from the tumor detection likelihood map was represented by the maximum confidence value within that lesion candidate, being considered a true positive if the Dice similarity coefficient with the ground truth was at least 0.1. 

To compare the three different PDAC-detection configurations, the ten trained models for each were applied individually to the test set. A permutation test with 100,000 iterations was then used to assess statistically significant differences between the area under the ROC curve (AUC-ROC) and partial area under the FROC curve (pAUC-FROC), which was calculated in the interval of [0.001-5] false positives per patient. A confidence level of 97.5\% was used to assess statistical significance (with Bonferroni correction for multiple comparisons).

The final performance for each configuration was obtained by ensambling the predictions of the ten models.

\begin{figure}[H]
\captionsetup{justification=centering}
\widefigure
        \begin{subfigure}[b]{0.5\textwidth}
                \includegraphics[width=\linewidth]{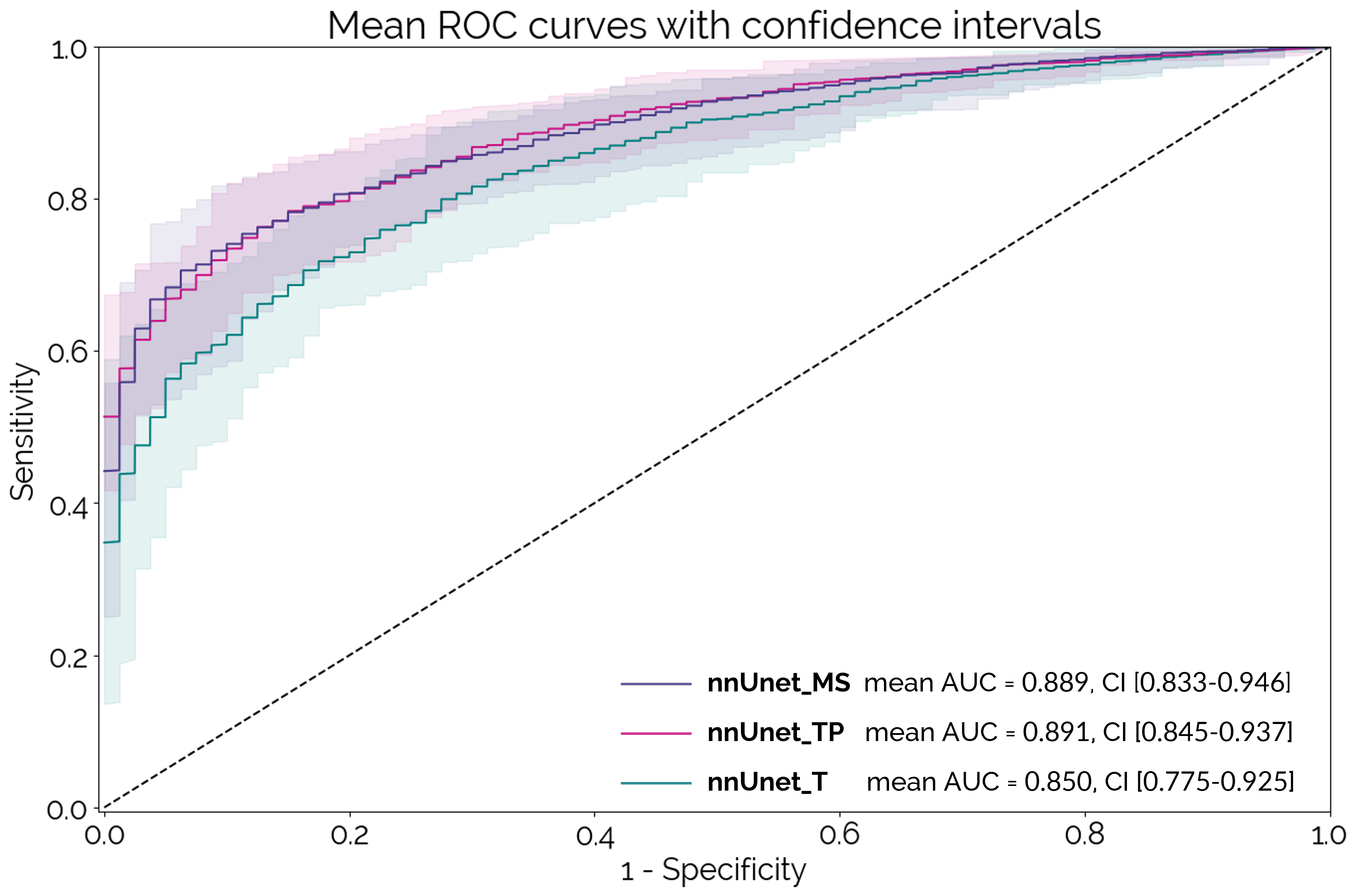}
                \label{fig:mean_ROC_test}
        \end{subfigure}%
        \begin{subfigure}[b]{0.5\textwidth}
                \includegraphics[width=\linewidth]{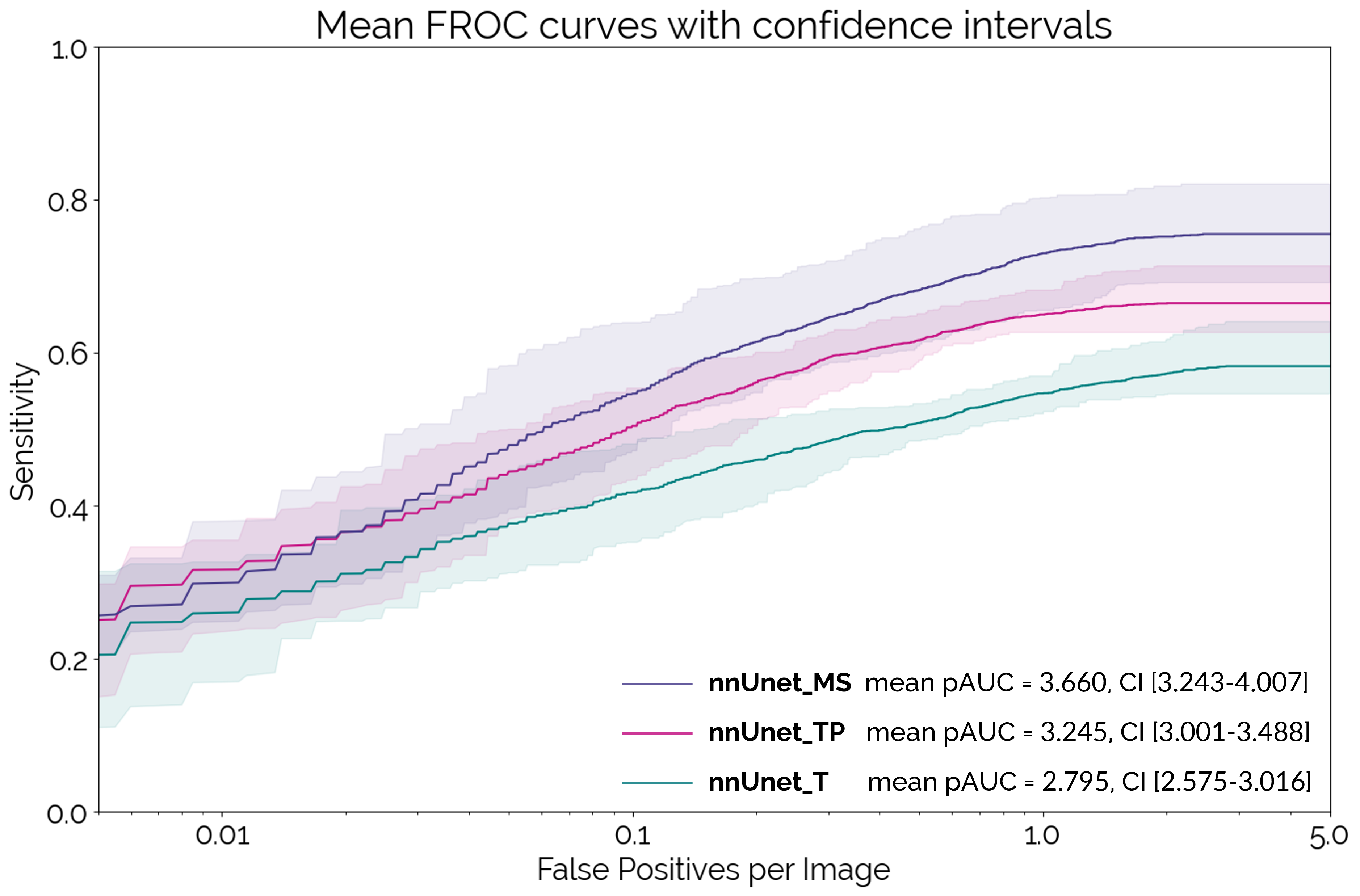}
                \label{fig:mean_FROC_test}
        \end{subfigure}
        \caption{Mean ROC and FROC curves with respective confidence intervals for the external test set.}\label{fig:test_results}
\end{figure}

\begin{figure}[H]	
\widefigure
\captionsetup{justification=centering}

        \begin{subfigure}[b]{0.5\textwidth}
                \includegraphics[width=\linewidth]{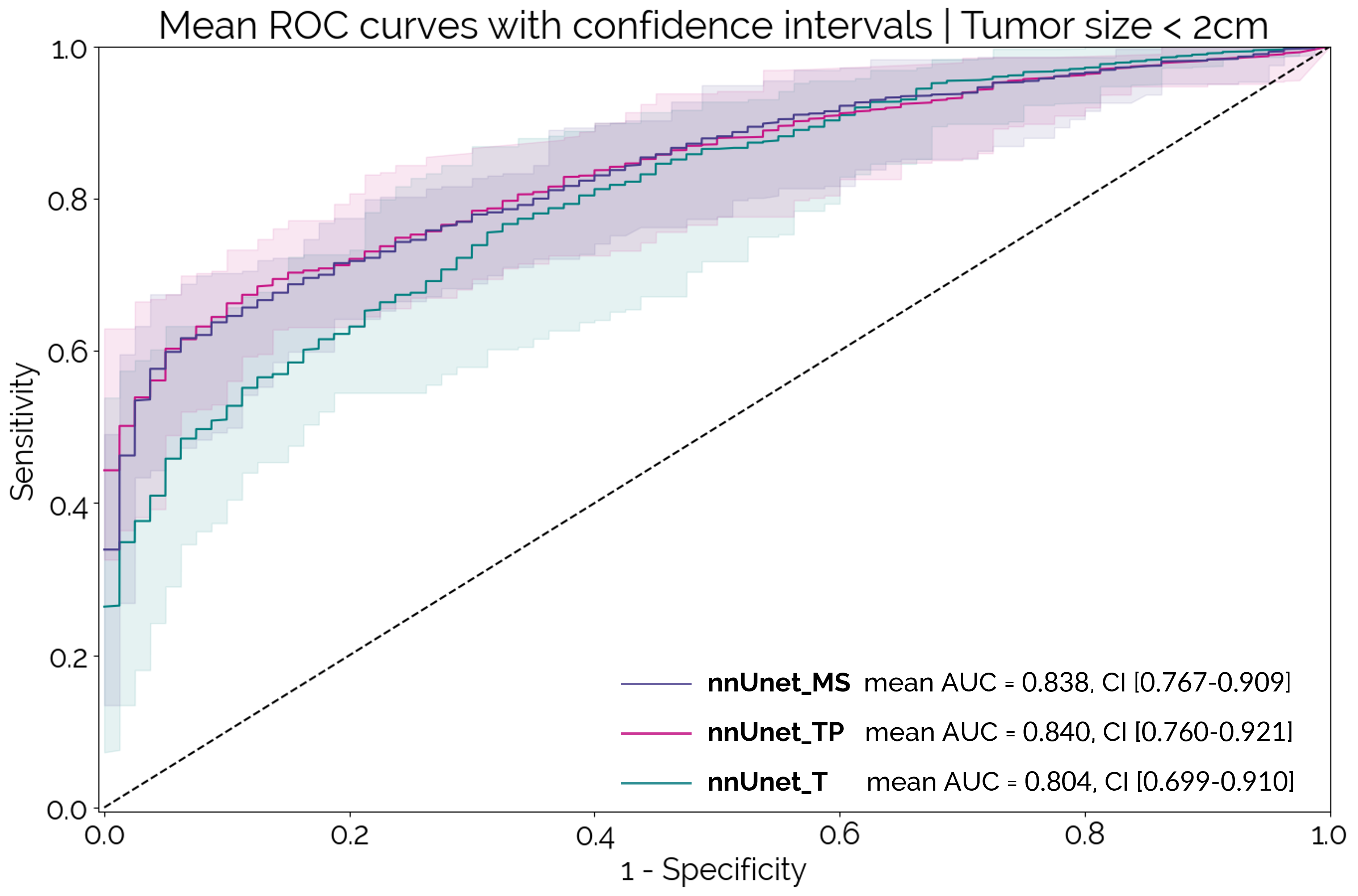}
                \label{fig:mean_ROC_validation}
        \end{subfigure}%
        \begin{subfigure}[b]{0.5\textwidth}
                \includegraphics[width=\linewidth]{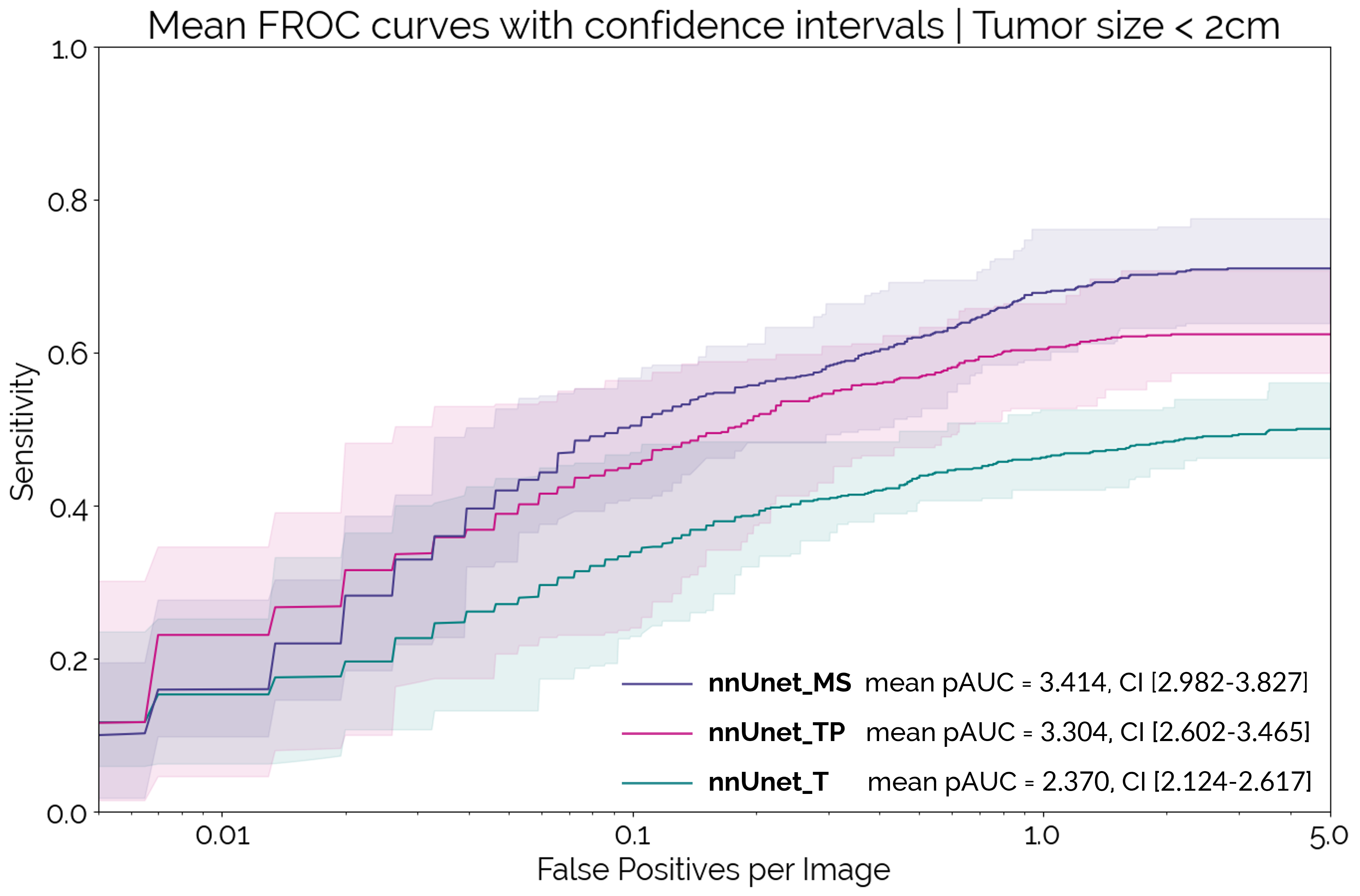}
                \label{fig:mean_FROC_validation}
        \end{subfigure}
        \caption{Mean ROC and FROC curves with respective confidence intervals for the external set considering only the subgroup of tumors with size < 2cm.}\label{fig:test_results_subgroup}
\end{figure}

\end{paracol}
\nointerlineskip
% Example of a figure that spans the whole page width (the commands \widefigure and \begin{paracol}{2}, \linenumbers, and\switchcolumn need to be present). The same concept works for tables, too.
\begin{paracol}{2}
\switchcolumn

%%%%%%%%%%%%%%%%%%%%%%%%%%%%%%%%%%%%%%%%%%
\section{Results}
The clinical characteristics of the patients in the PDAC cohort are summarized in Table~\ref{tabel:patient_charachteristics}. For the non-PDAC cohort, the mean age was 52.3\textpm21.4 (years), and there were 54 female and 69 male patients.

The performance of the three different PDAC detection network configurations on the internal 5-fold cross validation sets are shown in Table~\ref{tabel:validation_table}. At the patient level, the \textit{nnUnet\_MS} achieves the best performance, with a AUC-ROC of 0.991. Regarding lesion localization performance, the three configurations achieve similar pAUC-FROC, with the \textit{nnUnet\_MS} and \textit{nnUnet\_TP} performing slightly better than the \textit{nnUnet\_T}. 

% The MDPI table float is called specialtable
\begin{specialtable}[H] 
\centering
\captionsetup{justification=centering}
\caption{Clinical characteristics of the patients in the PDAC cohort. Data are mean\textpm standard deviation or median (interquartile  range). The tumor stages are: I-locally resectable, II-borderline resectable, III-locally advanced, IV-metastasized. \label{tabel:patient_charachteristics}}
%%% \tablesize{} %% You can specify the fontsize here, e.g., \tablesize{\footnotesize}. If commented out \small will be used.
\begin{tabular}{cc}
\midrule
Age (years)  	            & 69.2\textpm  8.5 \\
Gender (M/F)		        & 67/52\\
Tumor Stage (I/II/III/IV)	& 22/21/47/29 \\
Tumor size (cm) 	        & 2.8 (2.3-3.7)\\
\bottomrule
\end{tabular}
\end{specialtable}

\begin{specialtable}[H] 
\centering
\captionsetup{justification=centering}
\caption{Internal 5-fold cross-validation results for each configuration.\label{tabel:validation_table}}
%%% \tablesize{} %% You can specify the fontsize here, e.g., \tablesize{\footnotesize}. If commented out \small will be used.
\begin{tabular*}{\linewidth}{@{\extracolsep{\fill}} ccc}
\toprule
Configuration	& mean AUC-ROC (95\%CI)	& mean pAUC-fROC (95\%CI)\\
\midrule
\textbf{nnUnet\_T}		& 0.963	(0.914-1.0)		& 3.855 (3.156-4.553)\\
\textbf{nnUnet\_TP}		& 0.986	(0.956-1.0)		& 3.999 (3.252-4.747)\\
\textbf{nnUnet\_MS}		& 0.991	(0.970-1.0)		& 3.996 (3.027-4.965)\\

\bottomrule
\end{tabular*}
\end{specialtable}

The mean ROC and FROC curves obtained on the external test set with each PDAC detection network configuration are shown in Figure~\ref{fig:test_results}, with the respective 95\% confidence intervals. These curves were calculated using the 10 different trained models (2 initialisations with 5 fold cross-validation) for each configuration.  The \textit{nnUnet\_MS} and \textit{nnUnet\_TP} both achieve AUC-ROC around 0.89, which is significantly higher than the \textit{nnUnet\_T} ($p=0.007$ and $p=0.009$, respectively). At a lesion level, the \textit{nnUnet\_MS} achieves a significantly higher pAUC-FROC than both the \textit{nnUnet\_TP} and \textit{nnUnet\_T} ($p<10^{-4}$).

\end{paracol}
\nointerlineskip
% Example of a figure that spans the whole page width (the commands \widefigure and \begin{paracol}{2}, \linenumbers, and\switchcolumn need to be present). The same concept works for tables, too.
\begin{paracol}{2}
\switchcolumn

There were 73 tumors with size < 2 cm in the MSD dataset. Figure~\ref{fig:test_results_subgroup} shows the patient and lesion level results for each configuration on this sub-set of smaller tumors. In a patient level, the AUC-ROC decreases in about 0.05 for each configuration, when compared to the results obtained on the whole dataset. The \textit{nnUnet\_MS} and \textit{nnUnet\_TP} continued to outperform the \textit{nnUnet\_T}, although the differences were not statistically significant at a confidence level of 97.5\% ($p=0.034$ and $p=0.077$ respectively). Regarding lesion-level performance, the pAUC-FROC for the \textit{nnUnet\_MS} was still significantly higher than for the \textit{nnUnet\_TP} and \textit{nnUnet\_T} ($p<10^{-4}$ and $p=4.8\cdot 10^{-4}$ respectively).

The results obtained by ensembling the 10 models for each configuration are shown in Table~\ref{tabel:enamble_results}.

\end{paracol}
\nointerlineskip
% Example of a figure that spans the whole page width (the commands \widefigure and \begin{paracol}{2}, \linenumbers, and\switchcolumn need to be present). The same concept works for tables, too.
\begin{paracol}{2}
\switchcolumn

\begin{specialtable}[H] 
\centering
\captionsetup{justification=centering}
\caption{Ensemble results for each configuration on the whole test set and the subgroup of tumors with size < 2 cm.\label{tabel:enamble_results}}
%%% \tablesize{} %% You can specify the fontsize here, e.g., \tablesize{\footnotesize}. If commented out \small will be used.
\begin{tabular*}{\linewidth}{@{\extracolsep{\fill}} cccc}
\toprule 
Subgroup & Configuration	& AUC-ROC & pAUC-FROC \\
\midrule
\multirow{3}{*}{Whole Test Dataset} & \textbf{nnUnet\_T}		& 0.872		& 3.031\\
  & \textbf{nnUnet\_TP}		& 0.914	& 3.397\\
 & \textbf{nnUnet\_MS}		& 0.909 & 3.700\\
\midrule
\multirow{3}{*}{Tumors size < 2cm} & \textbf{nnUnet\_T}		& 0.831 & 2.671\\
  &\textbf{nnUnet\_TP}		& 0.867	& 3.289\\
  &\textbf{nnUnet\_MS}		& 0.876	& 3.553\\

\bottomrule
\end{tabular*}
\end{specialtable}

Figure~\ref{fig:output_example} shows an example of the network outputs of \textit{nnUnet\_TP} and \textit{nnUnet\_MS} for an isoattenuating lesion in the neck-body of the pancreas. This lesion was missed by both the \textit{nnUnet\_T} and \textit{nnUnet\_TP}, but could be correctly identified by the \textit{nnUnet\_MS} model. 

\begin{figure}[H]
\widefigure
\includegraphics[width=\linewidth]{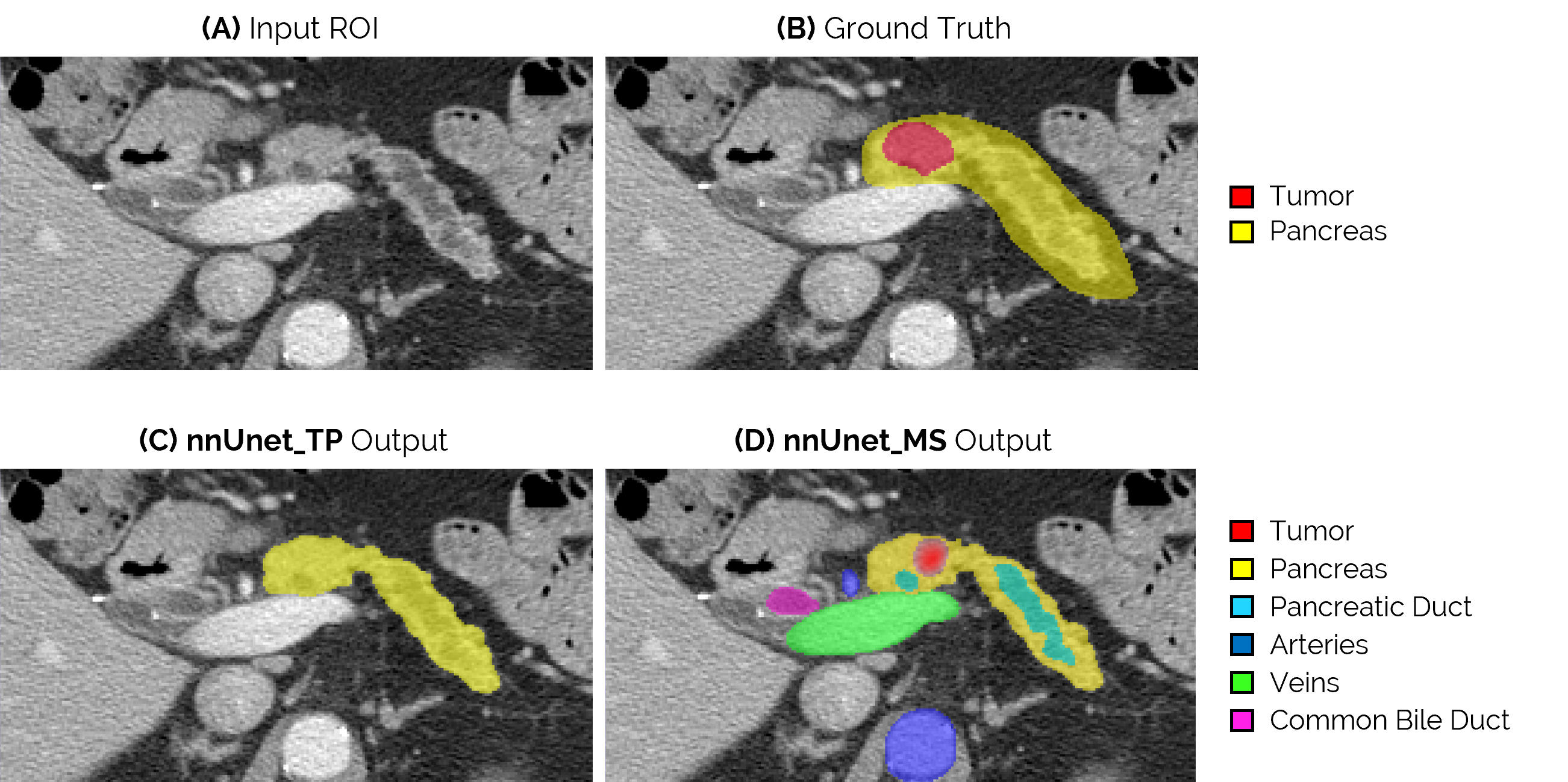}
\label{fig:output_example}
\caption{Example of an isoattenuating  tumor from the external test set which was missed by both the \textit{nnUnet\_T} and \textit{nnUnet\_TP} but could be correctly localized by the \textit{nnUnet\_MS}. (A) slice of the original ROI input; (B) ground truth segmentation of tumor and pancreas; (C) output of the \textit{nnUnet\_TP}, which in this case is only the pancreas segmentation as the tumor is not detected; (D) output of the \textit{nnUnet\_MS}, which is the segmentation of the detected tumor and surrounding anatomy.}\label{fig:output_example}
\end{figure}

\section{Discussion}

In this study, the state-of-the-art, self-configuring framework for medical segmentation \textit{nnUnet} \cite{Isensee2021NnU-Net:Segmentation} was used to develop a fully automatic pipeline for the detection and localization of PDAC tumors on CE-CT scans. Furthermore, the impact of integrating surrounding anatomy was assessed.

 A significant challenge of applying deep learning to PDAC detection is that the pancreas occupies only a small portion of abdominal CE-CT scans, with the lesions being an even smaller target within that region. Training and testing the networks with full CE-CT scans would be very resource consuming and provide a lot of unnecessary information regarding surrounding organs, distracting the model's attention from the pancreatic lesion location. In this way, it is necessary to select a small volume of interest around the pancreas, but having expert professionals manually annotate the pancreas before running each image through the network requires extra time and resources, which would significantly diminish the model's clinical usefulness. To address this issue, the first step in our PDAC detection framework is to automatically extract a smaller volume of interest from the full input CE-CT scan by obtaining a coarse pancreas segmentation with a low-resolution \textit{nnUnet}. To the best of our knowledge, this is the first study to develop a deep-learning-based fully automatic PDAC detection framework and externally validate it on a publicly available test set.

Previous studies have employed deep CNNs for automatic PDAC detection on CT scans \cite{Zhu2019Multi-scaleAdenocarcinoma, Xia2020DetectingEnsemble, Ma2020ConstructionDiagnosis, Liu2020DeepValidation, K2021FullyTumors, Wang2021LearningPrediction}, but only two studies validated their models on an external test set \cite{Liu2020DeepValidation, K2021FullyTumors}, with one using the publicly available pancreas dataset. Liu and Wu, et al. \cite{Liu2020DeepValidation} developed a 2D, patch-based deep learning model using the VGG architecture to distinguish pancreatic cancer tissue from non-cancerous pancreatic tissue. This approach required the prior expert delineation of the pancreas, which was then processed by the network in patches that were classified as cancerous or non-cancerous. At a patient level, the presence of tumor was then determined based on the proportion of patches that the model classified as cancerous. The authors tested this model on the external test set and achieved a AUC-ROC of 0.750 (95\%CI [0.749-0.752]) for the patch-based classifier, and 0.920 (95\%CI [0.891-0.948]) for the patient-based classifier \cite{Liu2020DeepValidation}. On the sub-group of tumors with size < 2 cm, the model achieved a sensitivity of 0.631 (0.502 to 0.747). More recently, Si, et al. \cite{K2021FullyTumors} developed an end-to-end diagnosis pipeline for pancreatic malignancies, achieving an AUC-ROC of 0.871 in an external test set, but validation on the public available dataset was not performed.

Our proposed automatic PDAC detection framework achieved a maximum ROC-AUC of 0.914 for the whole external test set and 0.876 for the subgroup of tumors with size < 2 cm. This performance is comparable to the current state-of-the-art for this test dataset \cite{Liu2020DeepValidation}, but with the advantage of being obtained automatically from the input image, with no user interaction required. Another advantage of our framework is that the lesion location is also identified and so the classification outcomes are immediately interpretable, since they directly arise from the network's segmentation of the tumor. Moreover, the achieved results set a new baseline performance for fully automatic PDAC detection, noticeably improving on the previous best AUC-ROC of 0.871 reported by Si, et al.

To the best of our knowledge, this is the first study to assess the impact of multiple surrounding anatomical structures in the performance of deep learning models for PDAC detection. Pancreatic lesions often present low contrast and poorly defined margins on CE-CT scans, with 5.4-14\% of tumors being completely iso-attenuating and impossible to differentiate from normal pancreatic tissue \cite{Blouhos2015TheAnalysis}. These iso-attenuating tumors are identified only by the presence of secondary imaging findings (such as the dilation of the pancreatic duct) and are more prevalent in early disease stages \cite{HoYoon2011Small, Blouhos2015TheAnalysis}.  In clinical practice, surrounding structures such as the pancreatic duct, the common bile duct, the surrounding veins (protomesenteric and splenic veins), and arteries (celiac trunk, superior mesenteric, common hepatic, and splenic arteries) are essential for PDAC diagnosis and local staging \cite{HoYoon2011Small, Blouhos2015TheAnalysis}. However, so far deep-learning models have focused only on the tumor and non-cancerous pancreas parenchyma, not taking the diagnostic information provided by all surrounding anatomy into account. 

 In this framework, the anatomy information was incorporated in the \textit{nnUnet\_MS} model, which was trained to segment not only the tumor and pancreas parenchyma but also several other relevant anatomical structures. The rationale behind this approach was that by learning to differentiate between the different types of tissue present in the pancreas volume of interest, the network could learn underlying relationships between the structures and consequently better localize the lesions. This network was compared to the \textit{nnUnet\_T}, which was trained to segment only the tumor, and the \textit{nnUnet\_TP}, trained to segment the tumor and pancreas parenchyma, in order to assess the impact of adding surrounding anatomy. 
 
 The results on the external test set show that, at a patient level, there is a clear benefit in adding the pancreas parenchyma when compared to training with only the tumor segmentation, as both the \textit{nnUnet\_TP} and \textit{nnUnet\_MS} achieved a significantly higher AUC-ROC than the \textit{nnUnet\_T}. There were however no differences in the performances of the \textit{nnUnet\_TP} and \textit{nnUnet\_MS} networks. Contrastingly, at a lesion-level, there was a clear separation between the three FROC curves both on the whole test set and on the subgroup of tumors with size<2 cm (Figures~\ref{fig:test_results},\ref{fig:test_results_subgroup}), with the \textit{nnUnet\_MS} achieving significantly higher pAUC-FROC than the two other configurations. This shows that the addition of surrounding anatomy improves the model's ability to localize PDAC lesions. Figure~\ref{fig:output_example} illustrates the advantage of anatomy integration in the case of an iso-dense lesion that is obstructing the pancreatic duct, causing its dilation. Both the \textit{nnUnet\_T} and \textit{nnUnet\_TP} models fail to identify this lesion, as there are no visible differences between the tumor and healthy pancreas parenchyma. However, the \textit{nnUnet\_MS} can accurately detect its location in the pancreatic neck-body following the termination of the dilated duct. By providing supervised training to segment the duct and other surrounding structures, the neural model can better focus on the remaining regions in the pancreas parenchyma, which may explain its ability to detect faint tumors. Furthermore, the multi structure segmentation provided by the \textit{nnUnet\_MS} presents useful information to the radiologist that can assist the interpretation of the network output regarding the tumor.

Despite the promising results, there are two main limitations to this study. First, the models were trained with a relatively low number of patients and only included tumors in the pancreatic head, which could be holding back the performance on external cohorts with heterogeneous imaging data. We are currently working on extending the training dataset to incorporate more patients, including tumors in the body and tale of the pancreas, in order to mitigate this issue. Second, training the anatomy segmentation network requires manual labeling of the different structures, which is resource-intensive. To address this problem, we only manually labeled the images from the PDAC-cohort and used self-learning to automatically segment the non-PDAC cohort, which could be introducing errors in training. Like the previous issue, the solution to this problem is to increase the size of the training dataset so that the model can learn better representations of the anatomy and consequently perform higher quality automatic annotations.

%%%%%%%%%%%%%%%%%%%%%%%%%%%%%%%%%%%%%%%%%%
\section{Conclusions}

This study proposes a fully automatic, deep-learning-based framework that can identify whether a patient suffers from PDAC or not and localize the tumor in CE-CT scans. The proposed models achieve a maximum AUC of 0.914 in the whole external test set and 0.876 for the subgroup of tumors with size < 2 cm, indicating that state of the art deep learning models are able to identify small PDAC lesions and could be useful at assisting radiologists in early PDAC diagnosis. Moreover, we show that adding surrounding anatomy information significantly increases model performance regarding lesion localization.
%%%%%%%%%%%%%%%%%%%%%%%%%%%%%%%%%%%%%%%%%%
\vspace{6pt} 

%%%%%%%%%%%%%%%%%%%%%%%%%%%%%%%%%%%%%%%%%%
%% optional
%\supplementary{The following are available online at \linksupplementary{s1}, Figure S1: title, Table S1: title, Video S1: title.}

% Only for the journal Methods and Protocols:
% If you wish to submit a video article, please do so with any other supplementary material.
% \supplementary{The following are available at \linksupplementary{s1}, Figure S1: title, Table S1: title, Video S1: title. A supporting video article is available at doi: link.} 

%%%%%%%%%%%%%%%%%%%%%%%%%%%%%%%%%%%%%%%%%%
\authorcontributions{Conceptualization, N.A., M.S., J.H and H.H.; methodology, N.A. and H.H.; software,  N.A. and J.B,.; validation,  N.A.; formal analysis,  N.A.; investigation,  N.A., M.S., G.L., J.B., J.H. and H.H.; resources, N.A., J.H and H.H.; data curation, N.A and G.L.; writing---original draft preparation, N.A.; writing---review and editing, M.S., G.L., J.B., J.H. and H.H.; visualization, N.A; supervision, J.H and H.H.; project administration, H.H.; funding acquisition, H.H. All authors have read and agreed to the published version of the manuscript.}

\funding{This project has received funding from the European Union’s Horizon 2020 research and innovation programme under grant agreement No 101016851, project PANCAIM.}

\dataavailability{The data presented in this study are available on request from the corresponding author dependent on ethics board approval. The data are not publicly available due to data protection legislation} 

\conflictsofinterest{The authors declare no conflict of interest.} 

%%%%%%%%%%%%%%%%%%%%%%%%%%%%%%%%%%%%%%%%%%
\end{paracol}
\reftitle{References}

% Please provide either the correct journal abbreviation (e.g. according to the “List of Title Word Abbreviations” http://www.issn.org/services/online-services/access-to-the-ltwa/) or the full name of the journal.
% Citations and References in Supplementary files are permitted provided that they also appear in the reference list here. 

%=====================================
% References, variant A: external bibliography
%=====================================
\externalbibliography{yes}
\bibliography{references.bib}

%%%%%%%%%%%%%%%%%%%%%%%%%%%%%%%%%%%%%%%%%%
%% for journal Sci
%\reviewreports{\\
%Reviewer 1 comments and authors’ response\\
%Reviewer 2 comments and authors’ response\\
%Reviewer 3 comments and authors’ response
%}
%%%%%%%%%%%%%%%%%%%%%%%%%%%%%%%%%%%%%%%%%%
\end{document}